\documentclass[twocolumn,showpacs, prl, preprintnumbers,superscriptaddress,amsmath,amssymb]{revtex4-1}
\usepackage{graphicx,color,bm,natbib, times, placeins} 

\begin{document}

\hfill {\small\it  J. Sun and A.E. Motter,  Phys. Rev. Lett.  \textbf{\textit{110}}, 208701 (2013)}

 \title{Controllability transition and nonlocality in network control} 

\author{Jie Sun}
\affiliation{Department of Mathematics and Computer Science, Clarkson University, Potsdam, NY 13699, USA}

\author{Adilson E. Motter}
\affiliation{Department of Physics and Astronomy, Northwestern University, Evanston, IL 60208, USA}


\begin{abstract}
\noindent
A common goal in the control of  a large network is to minimize the number of driver nodes or control inputs.  Yet, the physical determination of control signals and the  properties of the resulting control trajectories remain widely under-explored. Here we show that:  (i) numerical control fails in practice even for linear systems if  the controllability Gramian is {\it ill-conditioned}, which occurs frequently even when  existing controllability criteria are satisfied unambiguously; (ii) the control trajectories are generally {\it nonlocal} in the phase space, and their lengths are strongly anti-correlated  with the numerical success rate and number of  control inputs;  (iii) numerical success rate increases abruptly from zero to nearly one as the number of  control inputs is increased, a transformation we term {\it numerical controllability transition}. This reveals a trade-off between nonlocality of the control trajectory in the phase space and nonlocality of the control inputs in the network itself. The failure of numerical control cannot be  overcome in general by merely increasing numerical precision---successful control requires instead increasing the number of  control inputs beyond the numerical controllability transition.

\end{abstract}


\maketitle
A system is controllable if its state can be driven to different predefined states by a given set of input control signals, with controllability depending on both the number and the  placement of the control inputs \cite{sontag1991}. Control
often relies on the promise that the direct manipulation of relatively few degrees of freedom can render the entire system controllable. This promise has special meaning in the study of complex networks, where the large total number of nodes contrasts
with the limited number that can be directly controlled due to cost and physical constraints. The 
control of network systems 
is important 
in applications as diverse as
the operation of infrastructure networks \cite{infrast_contr}, 
coordination of moving sensors and robots \cite{bullo_2009}, 
devising of therapeutic interventions
\cite{albert_plos_2011}, management of ecological networks \cite{Lessard2005}, and  control of cascading failures in general \cite{Sahasrabudhe2011,network_review}, 
and it has received 
increased
attention in the recent physics literature~\cite{network_review,Lombardi2007, Porfiri2009, Kim2009,Sorrentino2009,Herrmann2012}.

A number of significant recent studies have focused on networks with $n$-dimensional linear time-invariant  
dynamics~\cite{Liu2011, Cowam2012,Yan2012,Nepusz2012}, 
    \vspace{-0.2cm}
\begin{equation}
    \vspace{-0.1cm}
\frac{dx(t)}{dt}=Ax(t)+Bu(t),
\label{eq1}
\end{equation}
with controllability usually determined by the Kalman's controllability matrix $K=\left[ B\; AB \cdots A^{n-1}B \right]$ \cite{Kalman1963}. 
Given the matrices $A$ and $B$,  control inputs $u(t)$ exist for any initial state $x^{(0)}$ and target state $x^{(1)}$ if and only if $K$ has full row rank.
In particular, if all nodes in a  network  have   intrinsic dynamics so that $A_{ii}\neq 0$ for all $i$, it follows that generically 
there exists $u(t)$ such that
the entire network can in theory be controlled by a single control input 
\cite{Cowam2012,Lin1974}.  

A fundamental question is, however, whether the control signals $u(t)$ can actually be constructed in practice. 
At first glance, this question may sound dull given that an explicit expression exists for $u(t)$ corresponding to the minimal-energy control trajectories in $t\in[t_0,t_1]$:
\begin{equation}
	u(t) = B^{T}\Phi^T(t_0,t)W^{-1}(t_0,t_1)\big[\Phi(t_0,t_1)x^{(1)} - x^{(0)}\big],
\label{eq2}
\end{equation}
where  $W(t_0,t_1) = \int_{t_0}^{t_1}\Phi(t_0,t) B B^T\Phi^T(t_0,t)dt$ is the controllability Gramian and $\Phi(t',t)=e^{(t'-t)A}$  \cite{Rugh1993}.
Incidentally, matrix $W(t_0,t_1)$ being invertible [hence Eq.~(\ref{eq2}) being well-defined]  is  equivalent to the commonly used Kalman's rank condition mentioned above. 
Quite surprisingly, despite this explicit solution and formal equivalence, we show that the determination of $u(t)$ 
is fundamentally limited in networks with more than a handful of nodes. This calls for a careful re-interpretation of existing controllability criteria.

Specifically, in this Letter we show that control fails in practice if the controllability Gramian is ill-conditioned,  
which can occur even when the corresponding Kalman's controllability matrix is well-conditioned. 
  We also show that  the control trajectory
from an initial to a target state  is generally nonlocal in the phase space
and remains finite-size  (in fact very long) even when the target state is brought arbitrarily close to the initial one.
The length  $\int_{t_0}^{t_1} \| \dot{x}(t) \| dt$ of such a trajectory generally increases with the condition number of the Gramian.
Both the nonlocality of the control trajectory and control failure rate are reduced by increasing the 
number of  control inputs. 
The latter manifests itself as a sharp transition as a function of the number of  control inputs, 
below which numerical control always fails and above which it succeeds. 
Aside from its implications for control, the characterization of such a numerical controllability 
transition adds a new  dimension to the research on structural \cite{Mendes2008}, 
dynamical \cite{Arenas2008}, and algorithmic complexity \cite{Moore2011} transitions in networks, which 
has had broad impacts \cite{book1,book2,book3},
with recent applications ranging from synchronization and percolation to epidemic processes \cite{eper1, eper2, eper3,eper4,eper5}.

There are known factors that can cause control to fail, including nonlinearity of the dynamics, parameter uncertainty, and stochasticity. Our results show that even in the most favorable case, in which the system is deterministic, autonomous,  and linear, the disparity between theory and practice poses a fundamental limit on our ability to control large networks.

\begin{figure}[t!]
  \begin{center}
    \includegraphics[width=.48\textwidth ]{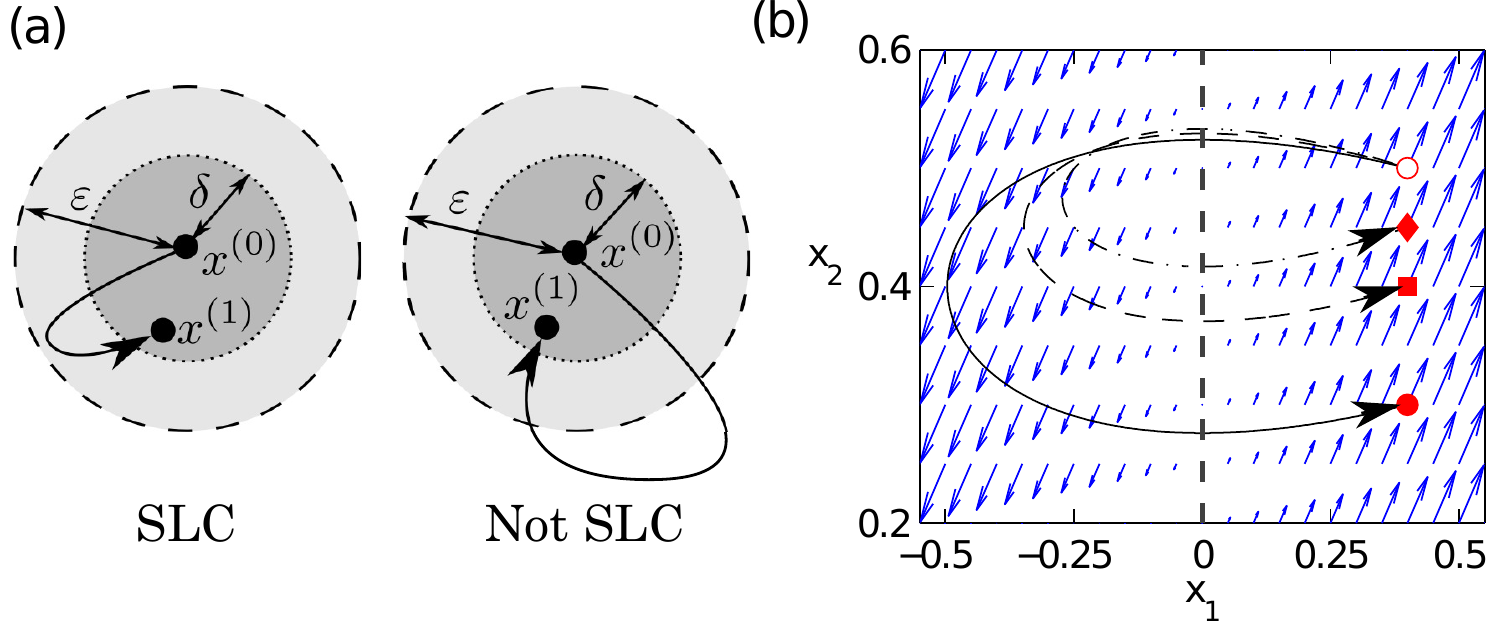}
        \vspace{-0.8cm}
  \end{center}
  \caption{(a) Illustration of a state that is SLC (left) and of a state that is not  (right).
  (b) Example  system $\dot{x}_1 = x_1 + u_1(t)$, $\dot{x}_2 = x_1$,  
  where any state not on the line $x_1=0$ is not SLC; the curves indicate minimal-energy  control trajectories for the given initial state (open symbol) and target states (solid symbols).  
  The background arrows indicate the vector field in the absence of control. As this two-dimensional example suggests, almost all states of linear systems described by Eq.~(\ref{eq1}) 
 are not SLC 
 whenever the number of control inputs is smaller than the number of dynamical variables. 
     \vspace{-0.3cm}
  }
  \label{fig1}
\end{figure}

We first establish the nonlocality of control trajectories. We say that a state $x^{(0)}$ of system (\ref{eq1}) is strictly locally controllable (SLC) if for any ball $B(x^{(0)}, \varepsilon)$ of radius $\varepsilon>0$ centered at $x^{(0)}$  there is a radius $\delta>0$ such that any target $x^{(1)}$  inside the concentric ball $B(x^{(0)},\delta)$  can be reached from $x^{(0)}$ with a control trajectory entirely inside  $B(x^{(0)},\varepsilon)$---see Fig.~\ref{fig1}(a). 
Note that a state can be locally controllable, in the sense that control trajectories always exist for neighboring target states, and yet not be SLC. Figure~\ref{fig1}(b) shows one such example in two dimensions:  for a state in the   $x_1>0$ half-plane,  the control trajectories to any neighboring target state with a smaller $x_2$-component  necessarily cross into the  $x_1<0$ half-plane (symmetric conclusions hold for initial states in the other half-plane).

We show that this result is in fact general for any controllable system in which one or more components are not directly controlled. Indeed, if the $k^{th}$ component  is not directly controlled, an initial state in the half-space $(Ax)_k>0$  can only be driven to a neighboring target state with $x^{(1)}_k<x^{(0)}_k$ if the control trajectory crosses into the half-space $(Ax)_k<0$, since otherwise $\dot{x}_k(t)=\big(Ax(t)\big)_k$ would be nonnegative and $x_k(t)$ would never decrease (analogous argument applies to the other half-space). Therefore, all states outside the hyperplane $(Ax)_k=0$ are not SLC, and hence almost all states are not SLC. 
The origin, on the other hand, is the only state that, when  $A$ is nonsingular, is SLC for any control matrix $B$ for which the system is controllable. 
The control trajectories defined by Eqs.~(\ref{eq1})-(\ref{eq2}), which minimize the energy $\int_{t_0}^{t_1}\|u(t)\|^2dt$, are given by
    \vspace{-0.1cm}
\begin{equation}
    \vspace{-0.1cm}
x(t) = \Phi(t,t_0)[x^{(0)}+M_{t_0,t_1,t}\big(\Phi(t_0,t_1)x^{(1)} - x^{(0)}\big)],
\label{eq3}
\end{equation}
where $M_{t_0,t_1,t}= W(t_0,t)W^{-1}(t_0,t_1)$ \cite{Rugh1993}.\  
The SLC property of the origin then follows from taking the norm on both sides of Eq.~(\ref{eq3}) for $x^{(0)}=0$ and upper-bounding the norm of the integral (exponential) terms by the integral (exponential) of the norm, 
which leads to
to $\|x(t)\|\leq C\|x^{(1)}\|$
for
$C=(t_1-t_0)\|BB^T\|\cdot\|W^{-1}(t_0,t_1)\|$. 
What sets the origin aside is that, for invertible $A$,  it corresponds to the only fixed point of the system.  
Having established the nonlocality of control trajectories for typical states in general, we now study in detail the minimal-energy control trajectories. 

We focus on undirected connected networks endowed with the dynamics in Eq.~(\ref{eq1}) for matrices $A$ with nonzero diagonal elements. These networks generically satisfy the Kalman's controllability condition for a single control input.
We consider  Erd\H{o}s-R\'enyi  (ER) networks
for a given number of nodes $n$ and
connection probability $p$. 
We address the impact of network structure by also considering networks generated using the configuration model \cite{book2} 
for scale-free (SF) degree 
distributions $P(k)\sim k^{-\beta}$ for $k\ge k_{\min}$, where in our simulations the minimum degree $k_{\min}$ is chosen to keep the average degree fixed as $\beta$ is  varied.
The edges and the diagonal elements $A_{ii}$  are assigned weights drawn from a uniform distribution in $[-1,1]$.  
Without loss of generality, 
we assume that the nodes of the networks are one-dimensional dynamical systems [i.e., $x(t)=(x_1(t), x_2(t),..., x_n(t))$, where $x_i(t)$ is a scalar variable
representing the state of node $i$]
and that the control matrix $B$ is diagonal upon row permutation, so that there is a one-to-one relationship between control inputs
and driver nodes. 
For each network, number $q$ of driver nodes, and given initial and target states,  we calculate numerically the minimal-energy control trajectories given by Eq.~(\ref{eq3}). 
Unless noted otherwise, we choose the driver nodes randomly at each independent realization
and consider the control time window $t_0\le t \le t_1$ for $t_0=0$ and $t_1=0$. 
Numerical control is declared successful if the numerically calculated state $x(t_1)$ is within a given distance $\eta\ll\|x^{(1)}-x^{(0)}\|$ of the 
target state $x^{(1)}$, where in our simulations we use $\eta=10^{-6}$.

\begin{figure}[t!]
  \begin{center}
    \includegraphics[width=.48\textwidth ]{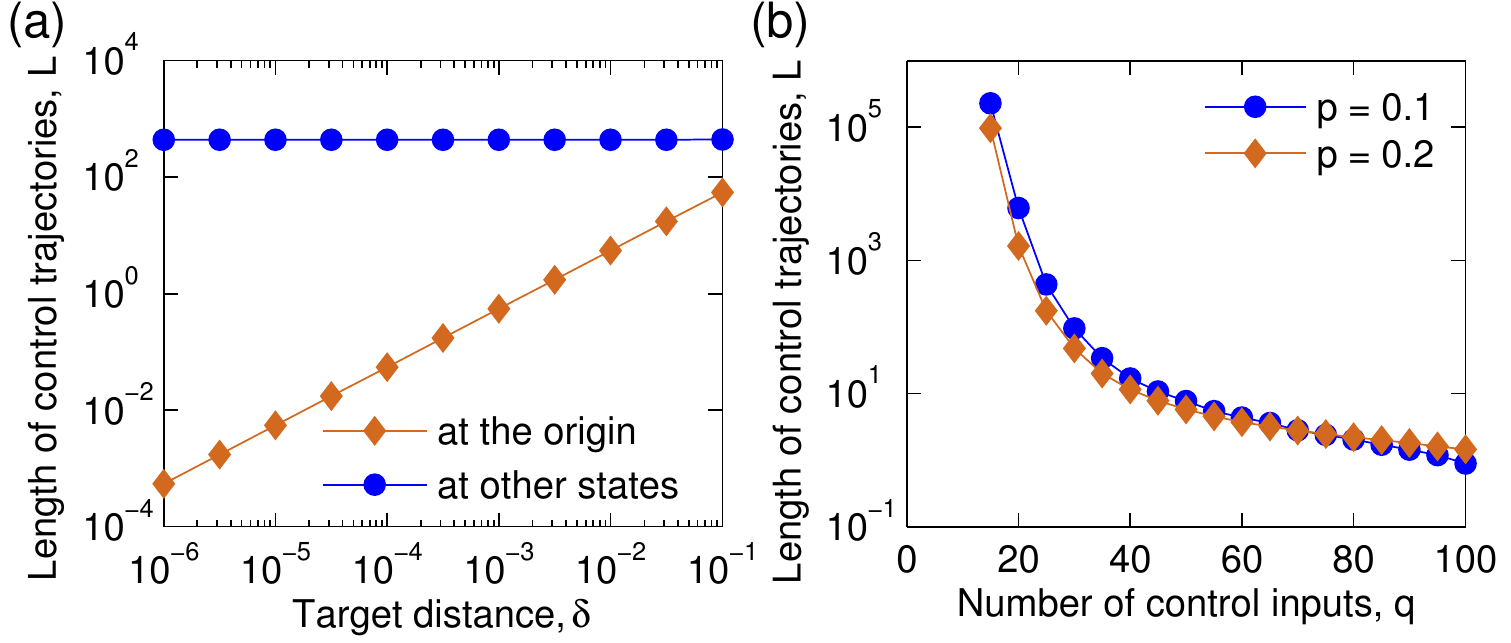}
        \vspace{-0.7cm}
  \end{center}
  \caption{(a) Average length of the control trajectories as a function of the distance from the target when the initial states are away from the origin versus when they are at the origin.  In the former case the initial states were chosen randomly on the unit sphere centered at the origin. In both cases  the target states are randomly oriented $\delta$ apart and $q=25$.
  (b) Average length of the control trajectories as a function of the number of control inputs for initial states away from the origin and $\delta=10^{-2}$.
Each data point corresponds to $1,000$ realizations, for ER networks with $n=100$. 
    \vspace{-0.3cm}
  }
  \label{fig2}
\end{figure}

Figure \ref{fig2} shows the average length $L$ of the control trajectories. For typical initial states, $L$  does not approach zero (and in fact remains essentially constant) as the distance to the targets is reduced, in stark contrast with the case in which the initial state is at the origin [Fig.~\ref{fig2}(a)].  This behavior becomes more pronounced  when the number of driver nodes is small, reaching $L=10^5$ for
 $q/n=0.15$;
conversely,  $L$ decreases as $q$ is increased [Fig.~\ref{fig2}(b)]. The latter can be understood in terms of our analytical argument above, since the smaller $n-q$ the fewer hyperplanes $(Ax)_k=0$ the control trajectory has to cross in order for  $\dot{x}_k(t)$ to acquire the right sign for every component $k$ not directly controlled.

\begin{figure}[t]
  \begin{center}
    \includegraphics[width=.48\textwidth ]{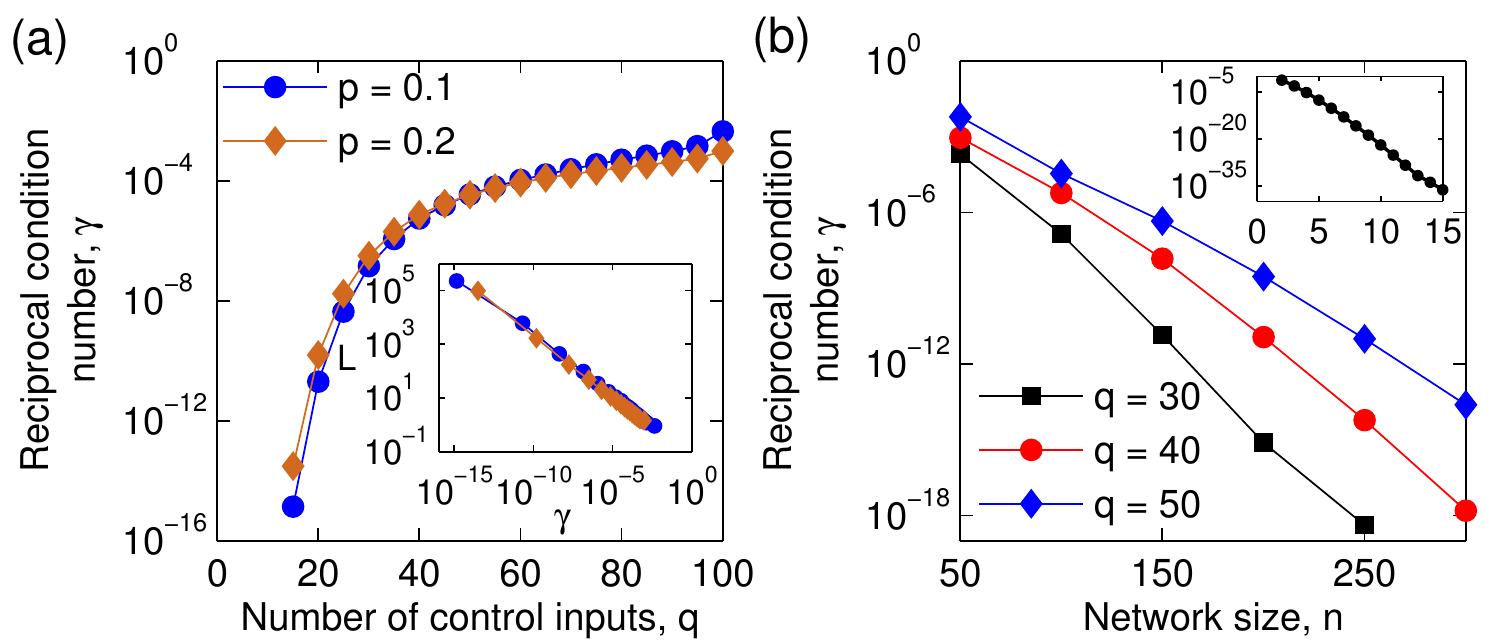}
        \vspace{-0.6cm}
  \end{center}
  \caption{Reciprocal condition number of the controllability Gramian $W$ as a function of (a) the number of control inputs for $n=100$ and (b) the network size for ER networks with $np = 10$.
Insets: relation between the average length of the control trajectories and the reciprocal condition number of $W$ [panel (a)]; reciprocal condition number of $W$ for the (analytically solvable) chain networks $C(n)$ with  $q=1$ (defined in the text) [panel (b)]. 
The statistics and parameters not shown are the same used in Fig.~\ref{fig2}(b).
    \vspace{-0.3cm}
  }
  \label{fig3}
\end{figure}

Given the mathematical equivalence between the Kalman's rank condition and the invertibility of the controllability Gramian, one might be tempted to assume that an ill-conditioned controllability Gramian $W(t_0,t_1)$ is a consequence of  an ill-conditioned Kalman's controllability matrix $K$. Here we disprove this conjecture by showing that the controllability Gramian can be nearly singular even when the corresponding controllability matrix is well-conditioned. This is best characterized by the reciprocal condition number $\gamma$. As a mathematically treatable example, we consider a directed linear chain $C(n)$ containing $n$ nodes and no self-loops: 
	$1\rightarrow{2}\rightarrow{3}\dots\rightarrow{n}$.  We examine the control of this network through the direct control of the root node, i.e., node $1$.  The control matrix is such that $B_{i}=\delta_{1,i}$ and, assuming for simplicity  that the network is unweighted, matrix $A$ is given by $A_{ij}=\delta_{i,j+1}$ for $i,j=1\dots n$. It follows from Eq.~(\ref{eq1}) that 
$K$  
is the $n\times n$ identity matrix. Therefore, $K$ has full rank, has reciprocal condition number $\gamma=1$, and is in fact the  best conditioned  of all matrices.
Now, consider 
$W(t_0,t_1)$. Taking for convenience $t_0=0$,  we can show that 
$W(0,t_1) = \int_{0}^{t_1}g(\tau)g^{T}(\tau)d\tau$,
where
$g(\tau) = \Big[1,-\tau,\frac{(-\tau)^2}{2!},\dots,\frac{(-\tau)^{i-1}}{(i-1)!},\dots,\frac{(-\tau)^{n-1}}{(n-1)!}\Big]^T$.
Thus, we can calculate the analytic expression for each entry of the Gramian:
$W(0,t_1)_{ij} = -\frac{(-t_1)^{i+j-1}}{(i+j-1)(i-1)!(j-1)!}$.
For fixed control time $t_1$, the reciprocal condition number $\gamma$  of $W(0,t_1)$ decreases exponentially as a function of the number of nodes in the linear chain,
as illustrated in the inset of Fig.~\ref{fig3}(b) for $t_1=1$. Therefore, as the size of the linear chain increases, the controllability Gramian quickly becomes nearly singular, 
making the numerical control of system (\ref{eq1}) nearly impossible,  even though the Kalman's controllability matrix remains well-conditioned.

Figure~\ref{fig3} shows that this behavior is indeed general for the ER networks we consider. The reciprocal condition number of $W$ decreases 
exponentially
as the number of control inputs is reduced [Fig.~\ref{fig3}(a)] or the size of the network is increased [Fig.~\ref{fig3}(b)], while the reciprocal condition number of $K$ (not shown) is 
generally 
larger than  $\sqrt{\gamma(W)}$ \cite{submatrix_K}.  Moreover, it follows that the average length of the control trajectories is strongly correlated with the condition number of the  controllability Gramian [Fig.~\ref{fig3}(a), inset].  
This can be rationalized by noting that both $L$ and $\gamma(W)$ are measures of how difficult it is to actually control the system in practice. 
Therefore, although the Kalman's controllability matrix  $K$ has attractive analytical properties, its use in practice requires caution.  In particular, since the minimum-energy control involves the inversion of the controllability Gramian $W$ (rather than of the Kalman's controllability matrix), we note that it is the  numerical rank 
 of $W$ that is often most relevant. 

A full-rank matrix is {\it numerically} rank deficient if one or more of its singular values fall below the predefined numerical threshold. 
If such a full-rank matrix has a bounded largest singular value, the numerical rank being smaller than the actual rank is necessarily related to a small reciprocal condition number $\gamma$ and vice versa, since $\gamma$ is the ratio of  the smallest to the largest singular value.  
While leading to a deficient numerical rank for matrix $W$  when $q/n$ is small, this relation is not a factor for  matrix $K$ because
this matrix
 is reasonably 
well-conditioned in the networks we consider, indicating that the numerical calculation of its rank is reliable. Indeed, for all networks considered in Fig.~\ref{fig3} we verified that the numerical rank of $K$ is $n$ for any numerical threshold smaller than $10^{-3}$.

\begin{figure}[t!]
  \begin{center}
    \includegraphics[width=.48\textwidth ]{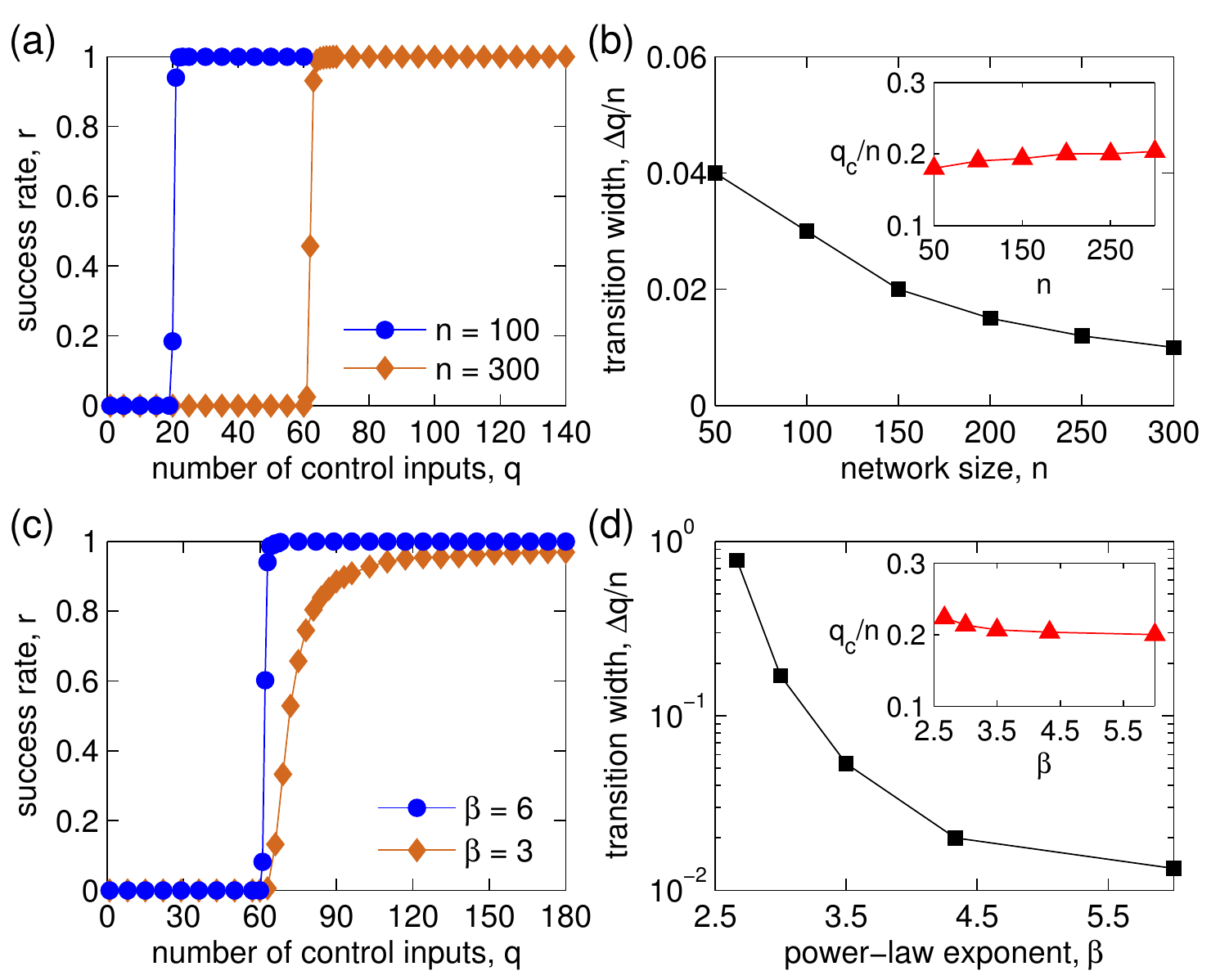}
    \vspace{-0.6cm}
  \end{center}
 \caption{Numerical controllability transitions for (a,b) ER networks of different sizes and  (c,d) SF networks with different scaling exponents. 
(a,c) Success rate as function of the number of control inputs.
(b,d) Transition width (main panel) and transition point (inset). 
The average degree was set to $10$ and, for the SF networks,  $n=300$.
The statistics and parameters not shown are the same used in Fig.~\ref{fig2}(b).
    \vspace{-0.4cm}
  }
  \label{fig4}
\end{figure}

Figure \ref{fig4}(a) shows that the numerical success rate increases sharply from 
zero to one 
as the number of driver nodes is increased.  According to the Kalman's rank condition, all networks we simulate are controllable in theory for $q$ as small as $1$.  
The transition in Fig.~\ref{fig4}(a) is a direct consequence of the decrease in the condition number of the controllability Gramian and the limitation it imposes on numerical calculations. 
The numerical error in computing  Eq.~(\ref{eq3}) is dominated by the round-off errors in the calculation of $W^{-1}$. Taking $t=t_1$ and using tilde to denote numerically computed  values, we obtain 
    \vspace{-0.1cm}
\begin{equation}
\|\tilde{x}^{(1)} -x^{(1)}  \| \lessapprox D \|W(\tilde{W}^{-1}-W^{-1})\|, 
\label{approx}
\end{equation}
where $z=\Phi(t_0,t_1)x^{(1)}-x^{(0)}$ and $D=\|\Phi(t_1,t_0)\|\cdot\|z\|$. 
If $W + \Delta W$ is the exact  inverse of $\tilde{W}^{-1}$, then 
the right side of Eq.~(\ref{approx}) is bounded from above by $D\| \Delta W \| \cdot \|\tilde{W}^{-1}\|  \approx D\|W\| \cdot \|W^{-1}\| \cdot  \frac{\| \Delta W \|}{\|W\|}$, where $\|W\| \cdot \|W^{-1}\| =1/\gamma(W)$ and $ \frac{\| \Delta W \|}{\|W\|}$ is of the order of the numerical precision $\epsilon$~\cite{Demmel1997}.
Thus, we predict that the transition to successful numerical control will occur in general as $\gamma(W)$ decreases past  
   \vspace{-0.2cm}
\begin{equation}
    \vspace{-0.2cm}
\gamma_c\approx D' \frac{\epsilon}{\eta} ~,  
\label{g_c}
\end{equation} 
where $D'$ is a constant determined by $t_1-t_0$, $x^{(0)}$, $x^{(1)}$ and $A$,
and $\eta$ is the radius of convergence. 
For double precision ($\epsilon\simeq 10^{-16}$) and $\eta=10^{-6}$, as considered in our simulations, 
the transition is predicted to occur around $\gamma=10^{-10}$ for $D'$ of order unity,
which agrees with our numerics.

To further characterize this transition, we analyze its width within the network ensemble, defined as $\Delta q/n \equiv (q_{c'}-q_c)/n$, where $q_c$ and  $q_{c'}$ mark the integer number of control inputs right below  $5\%$ and right above $95\%$ success rate, respectively. 
As shown in Fig.~\ref{fig4}(b), the transition becomes increasingly sharp as the size of the network increases. The transition point, which we take as being $q_c/n$ for the purpose of this discussion, 
is around $0.20$ and increases slowly as $n$ increases.

To address the impact of degree heterogeneity, we have also analyzed the controllability transition in SF networks. As shown in Fig.~\ref{fig4}(c,d), the transition becomes wider as the variance of the degree distribution increases  (i.e., $\beta$ decreases), indicating that  control fails more often the more heterogeneous the degree distribution (cf.\ Ref. \cite{Liu2011}), but the starting point of the transition is very insensitive to the degree distribution and is essentially the same for ER and SF networks with the same network size and average degree [e.g., Fig.~\ref{fig4}(a) and Fig.~\ref{fig4}(c) for $ n=300$].  Moreover, these conclusions do not depend sensitively on how the driver nodes are selected: we have verified that the curves in 
Fig.~\ref{fig4}(a,c) shift horizontally by less than the size of the symbols when the driver nodes are selected not randomly but instead as the lowest- or the highest-degree nodes in the network.

A few observations are in order.
First, one may ask whether the impossibility of controlling the system with a reduced number of driver nodes could be avoided by increasing the precision of the numerical calculations.  
Because the reciprocal condition number $\gamma(W)$ decreases exponentially as $q$ is reduced while $\gamma_c$ decreases only linearly as  $\epsilon$ is reduced, for large networks any realistic increase in precision will only have a limited effect in reducing the critical fraction of driver nodes  $q_c/n$ at which the numerical controllability transition takes place.
 Thus, tantamount to the impossibility of  long-term predictability in chaotic dynamics,  this fundamental limitation cannot be easily overcome  
 by increasing numerical precision and becomes even more pronounced in larger networks.

Second, subtle differences in the formulation of the dynamics in Eq.~(\ref{eq1}) have led to very different conclusions about the minimal number of driver nodes according to the Kalman's rank condition, being generically one if the diagonal elements of $A$ are nonzero  \cite{Lin1974,Cowam2012} and generally much larger than one if they are not \cite{Liu2011,Murota1987}. It is thus satisfying to find that in practice the results are far more robust against small changes in the model parameters.

Third, the nonlocality of  control trajectories identified here reveals an intriguing mechanism of failure in applying control results from linearized dynamics to their nonlinear counterparts \cite{sun2011}: such an approach fails because the control trajectories are required to go outside the neighborhood in which the linearization is valid.
Moreover, there are known examples of nonlinear systems that are globally controllable while their linearizations are not  \cite{Lynch2005},
and it is straightforward to identify network systems too which have this property \cite{comment}. It is thus natural to speculate that approaches that exploit nonlinear properties of the dynamics
(see, e.g., \cite{Khalil2001}), although potentially more involved, can be
 better suited to address nonlinear systems---for a general such approach specifically developed to control complex networks with nonlinear dynamics, see Ref.~\cite{sean2013}.

Our demonstration that either the control trajectory is nonlocal in the phase space or the control inputs are nonlocal in the network 
has several implications.  In practice, the former leads to failure of the numerical control and the latter points to an overhead that has to be accounted for in optimizing the number of driver nodes. For the random networks considered here, this gives rise to a sharp transition as a function of the number of driver nodes, below which numerical control always fails and above which it succeeds. These findings suggest the need of a controllability criterion that accounts not only for the existence but also for the actual computability of the control interventions: {\it the system is controllable in practice if and only if the controllability Gramian has full numerical rank}. 
We suggest that by applying such a criterion future research  may reveal a rich set of relations between controllability and network structure even when
no dependence is predicted by the Kalman's rank condition.

This study was supported by NSF grant DMS-1057128.


\begin{thebibliography}{99}
\bibitem{submatrix_K}
In the case of matrix $K$, 
the reported $\gamma$
was calculated from the submatrix formed by the first $n$ columns  of $K$, which suffices to determine its full row rank.
\end{thebibliography}
\end{document}